# Stability of E' centers induced by 4.7eV laser radiation in $SiO_2$

F. Messina*, M. Cannas

Dipartimento di Scienze Fisiche ed Astronomiche dell'Università di Palermo,Via Archirafi 36, I-90123 Palermo, Italy

## ABSTRACT

The kinetics of E' centers ($\equiv$Si•) induced by 4.7eV pulsed laser irradiation in dry fused silica was investigated by *in situ* optical absorption spectroscopy. The stability of the defects, conditioned by reaction with mobile hydrogen of radiolytic origin, is discussed and compared to results of similar experiments performed on wet fused silica. A portion of E' and hydrogen are most likely generated by laser-induced breaking of Si-H precursors, while an additional fraction of the paramagnetic centers arise from another formation mechanism. Both typologies of E' participate to the reaction with $H_2$ leading to the post-irradiation decay of the defects. This annealing process is slowed down on decreasing temperature and is frozen at T=200K, consistently with the diffusion properties of $H_2$ in silica.

* Corresponding author: F. Messina. Phone: +39 09162342218, Fax: +390 916162461
e-mail: fmessina@fisica.unipa.it

## 1. Introduction

The stability of point defects generated in amorphous silicon dioxide ($a$-SiO$_2$) by laser or ionizing radiation is often conditioned by their reactivity with mobile species such as molecular hydrogen [1-5]. This is particularly true for paramagnetic defects like the E' center, consisting in un unpaired electron spin on a silicon atom (≡Si•), whose generation is one of the main causes of degradation of optical transparency of $a$-SiO$_2$ in the ultraviolet (UV), due to its optical absorption (OA) band peaked at 5.8eV [3, 6-7] . Indeed, in presence of H$_2$, E' centers decay due to the following reaction [2-3,5,8-10]:

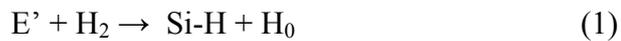

$$\text{E' + H}_2 \rightarrow \text{Si-H + H}_0 \qquad (1)$$

Reaction (1) is relevant for the applicative performance of the material since it permits a partial transparency recovery of irradiated silica glass. Most literature studies dealing with the generation of E' center have focused only on the stationary concentration of the defects measured after irradiation, whereas only a few have investigated its transient kinetics [10-13]; hence, the interplay between radiation-induced generation and decay due to (1) still needs to be completely clarified.

In recent experiments performed on wet fused silica samples, it was observed that 4.7eV laser irradiation at room temperature induces the generation of E' centers by breaking of Si-H precursors [10]. After the end of irradiation, the defects undergo a spontaneous decay whose kinetics lasts a few hours [5,10]. This process was attributed to reaction (1) by comparison of its kinetics with the diffusion parameters of H$_2$, and by the observation by electron spin resonance (ESR) of the concurrent growth of H(II) centers (=Ge-H•) [5], which may considered as a fingerprint of the presence of mobile hydrogen.

The aim of the present work is to clarify some still open issues on the stability of laser-induced E' centers in fused silica. In particular, the comparison between wet and dry materials is discussed, and the temperature dependence of the process at T<300K is analyzed to verify its consistency with reaction (1).

## 2. Materials and Methods

The experiments were performed on Infrasil301 dry (OH content ~10 ppm) fused silica samples, 5×5×2 mm$^3$ sized, produced by Heraeus Quartzglas by fusion of natural α-quartz powder in vacuum. The specimens were placed in high vacuum (10$^{-5}$mbar) inside a liquid He continuous flux cryostat, where they were irradiated perpendicularly to the minor surface, at 200K<T<300K, by the IV harmonic of the pulsed radiation emitted by a Q-switched Nd:YAG laser (hν≈4.7 eV, pulse width 5 ns, repetition rate 1Hz, energy density per pulse 40mJ/cm$^2$ measured by a pyroelectric detector).

OA spectra were measured *in situ* by a single-beam optical fiber spectrophotometer system (AVANTES S2000), working in the 200-400nm spectral range, and based on a $D_2$ lamp source and on a 2048 channel charge coupled diode (CCD) array detector. Measures were performed during the irradiation session, in the interpulse time range, and continued for a few 10$^3$s after switching off the laser.

ESR measurements on the signal of E' centers were performed by a spectrometer (Bruker EMX) working at a microwave frequency of 9.8GHz. The signal was detected with a 100KHz modulation frequency, a 0.01mT modulation amplitude and a 8×10$^{-4}$mW microwave power, low enough to prevent saturation. The E' centers were absent in as-grown samples as checked by preliminary measurements. Absolute concentration of the defects was calculated by comparison of the doubly-integrated signal with that of a reference sample where the absolute density was known from spin-echo measurements [14].

## 3. Results

An as-grown specimen ($S_1$) was irradiated at room temperature with $N_1=2.5\times10^3$ laser pulses. By monitoring *in situ* the difference OA spectra during and after the end of the irradiation session, we detected the 5.8eV band of E'-center, (inset of Fig. 1) whose intensity was found to depend on time. From the peak OA and the known absorption cross section of the defects [15] we calculated the concentration kinetics [E'](t), reported in Fig. 1. Just after exposure to $N_1$ pulses, the concentration is $[E']_M(N_1)=(3.0\pm0.3)\times10^{16}cm^{-3}$. As soon as the laser is switched off, we note a change of slope in the kinetics due to the begin of E' decay. [E'] continues to decrease for many hours, though with progressively decreasing rate. The most reliable estimate of the stationary asymptotic concentration $[E']_\infty$ of the defects is obtained by an ESR measurement performed a few days after irradiation, yielding $[E']_\infty(N_1)=(1.0\pm0.1)\times10^{16}cm^{-3}$. The post-irradiation kinetics of E' was measured with the same technique after different numbers ($N_2=6.6\times10^2$, $N_3=2.2\times10^2$) of laser pulses on other two as-grown specimens of the same material: results are reported in Fig. 1. The stationary concentrations were found to be $[E']_\infty(N_2)=(0.52\pm0.05)\times10^{16}cm^{-3}$, $[E']_\infty(N_3)=(0.23\pm0.02)\times10^{16}cm^{-3}$. We can define the ratio between stable and transient centers $\eta=[E']_\infty/([E']_M-[E']_\infty)$, which is found to depend on the number of pulses, being $\eta=(0.5\pm0.1)$, $(0.36\pm0.08)$, $(0.29\pm0.06)$ respectively for $N=N_1,N_2,N_3$.

We investigated also the effect of temperature on the post-irradiation stage of the kinetics, in the 200K<T<300K interval. In Fig. 2 we compare the decay of E' after exposure to $2\times10^3$ pulses at some representative temperatures. The curves are normalized to the concentrations at the end of exposure. The decay process becomes progressively slower on decreasing T, and is absent within experimental uncertainty at T=200K.

Finally, we performed a repeated irradiation experiment to find out if the post-irradiation kinetics is influenced by the previous history of the sample. In detail, when the decay was concluded and [E'] had reached its stationary value $[E']_\infty(N_1)$, the sample $S_1$ was irradiated a second time with 50 laser pulses. The concentration kinetics of E' during and after this second exposure is reported in

Fig. 3-A. For comparison, we report in Fig. 3-B the result of the same 50 pulses irradiation on an as-grown sample. The two kinetics are different; indeed, on the initially virgin sample, 50 pulses induce the generation of $[E']_M(50)=(0.36\pm0.04)\times10^{16}cm^{-3}$ defects, ~17% of which decay after 1 hour of the post-irradiation stage; at variance, re-irradiation of the $S_1$ sample induces a smaller variation of $[E']$ from $[E']_\infty(N_1)$: $(0.25\pm0.02)\times10^{16}cm^{-3}$; most important, the decay of the induced defects is faster (~55% after 1 hour).

## 4. Discussion

As discussed in the introduction, the post-irradiation effects observed in fused silica upon laser irradiation are due to diffusion and reaction of radiolytic $H_2$. In particular, in wet fused silica it was demonstrated that hydrogen and E' are generated together by breaking of Si-H precursors [10]:

$$Si\text{-}H + h\nu \rightarrow E' + H_0 \qquad (2)$$

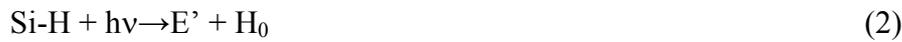

and $H_2$ is formed by dimerization of $H_0$ at the right side of (2). A consequence of (2) is that $[H_2]$ at the end of irradiation equals one half $[E']$: $[H_2]_M=1/2[E']_M$. Hence, the E' centers present at the end of exposure are almost completely erased by reaction (1) in the post-irradiation stage.

Present results (Fig. 1) confirm that also in dry fused silica E' centers are generated by laser radiation and decay in the post-irradiation stage by reaction (1). This attribution is also confirmed by the temperature dependence experiments. In fact, data in Fig. 2 demonstrate the post-irradiation decay to be a thermally activated process, frozen at $T_0$~200K. This finding is consistent with literature studies where diffusion of $H_2$ in silica has been characterized by the same threshold temperature [1,4,16] A more detailed quantitative analysis of the reaction kinetics is in progress with the aim to estimate its activation energy.

From comparison with wet materials, here too we can argue process (2) to be the most likely source of the hydrogen responsible for the decay of E'. This hypothesis is corroborated also by the following observation: Si-H and Si-OH groups are the two main forms in which bonded hydrogen may be accomodated in *a*-SiO$_2$. Now, the absence at the end of exposure of the 4.8eV band of non bridging oxygen hole center, ($\equiv$Si–O•) (see inset of Fig. 1), which would be observed if hydrogen was generated from breaking of Si–OH, strongly suggests Si-H as the most likely precursor for the mobile specie. In this respect, Infrasil materials are known to contain Si-H groups from Raman measurements on their typical signal at 2250cm$^{-1}$ [17].

However, the main difference between wet and dry materials is that in the latter the passivation of E' is incomplete, i.e. only a fraction of the initially induced defects disappear after the end of exposure. This result can be interpreted as follows: in dry glass a second formation channel of E' apart from (2) is active: as a consequence, concentration of generated E' is higher than ½[H$_2$], so that available H$_2$ is insufficient to completely cancel the induced defects. In this scheme, the concentration [E']$_M$ at the end of exposure consists in two portions: the defects [E']$_H$ coming from Si-H, plus a contribution [E']$_X$ arising from the second formation process. The above defined parameter η equals the ratio [E']$_X$/[E']$_H$ and depends on the number of pulses, this meaning that the two concurrent generation processes follow different growth curves. At the moment it is not possible to identify the generation process of the [E']$_X$, component, though the presence of oxygen deficient centers ($\equiv$Si-Si$\equiv$ vacancies) in dry fused glass suggests ionization of this precursor as a possible mechanism [3,7,18]. More studies may help to clarify this point.

It is not obvious *a priori* if reaction (1) involves all E' centers generated by irradiation or only the portion coming from the Si-H precursor. To clarify this issue, we start with writing down the reaction rate of (1) which, in the stationary state approximation for atomic hydrogen [19], is given by:

$$\frac{1}{2}\frac{d}{dt}[H_2] = \frac{d}{dt}[E'] = -2k[E'][H_2] \qquad (3)$$

where k is the reaction constant between $H_2$ and E'. If only the $E'_H$ centers were reactive with $H_2$, [E'] on the right side of (3) should be identified with $[E']_H$ and the decay kinetics would be independent from the previous history of the sample, not influenced by other contributions to the total E' concentration. Results in Fig. 3 disagree with this picture, as the decay of the defects induced by 50 pulses is accelerated when the sample, prior to irradiation, already hosts a concentration $[E']_\infty(N_1) \sim 10^{16} cm^{-3}$ due to a previous exposure. This is a consequence of the increase of the right side of (3) due to the addition of $[E']_\infty(N_1)$ to the total population of reacting centers. Then we conclude that all E' centers partecipate to the reaction of $H_2$ independently from their origin.

## 5. Conclusions

We studied by *in situ* OA spectroscopy the generation and isothermal annealing kinetics of E' centers induced in dry fused silica by 4.7eV laser irradiation. At temperatures higher than 200K, the induced defects undergo a partial post-irradiation decay due to reaction with mobile $H_2$. Hydrogen and a portion of E' centers are generated together from radiolysis of Si-H bonds, while also a second generation process contributes to the total concentration of the defects; both typologies of E' centers participate to the reaction with $H_2$.

**Acknowledgements**

The authors wish to thank Prof. R. Boscaino and group for support and enlightening discussions and Mr. Bodo Kühn (Heraeus Quarzglas GmbH) for providing useful information on materials. Technical assistance by G. Lapis and G. Napoli is also acknowledged.

**Figure Captions**

**Fig.1:** Growth kinetics of E' centers during 4.7eV pulsed laser irradiation and decay kinetics after 2500, 660, 220 laser pulses. Inset: difference absorption spectrum measured just after the end of the 2500 pulses irradiation.

**Fig.2:** Post-irradiation kinetics of E' after irradiations with 2000 laser pulses performed at three different temperatures. The curves are normalized to the concentration measured at the end of exposure.

**Fig.3:** Panel (B): Kinetics of [E'] induced by a 50 pulses irradiation on an as-grown sample. Panel (A): the same experiment, but performed on a sample which already contained $[E']\sim10^{16}cm^{-3}$ due to a previous irradiation.

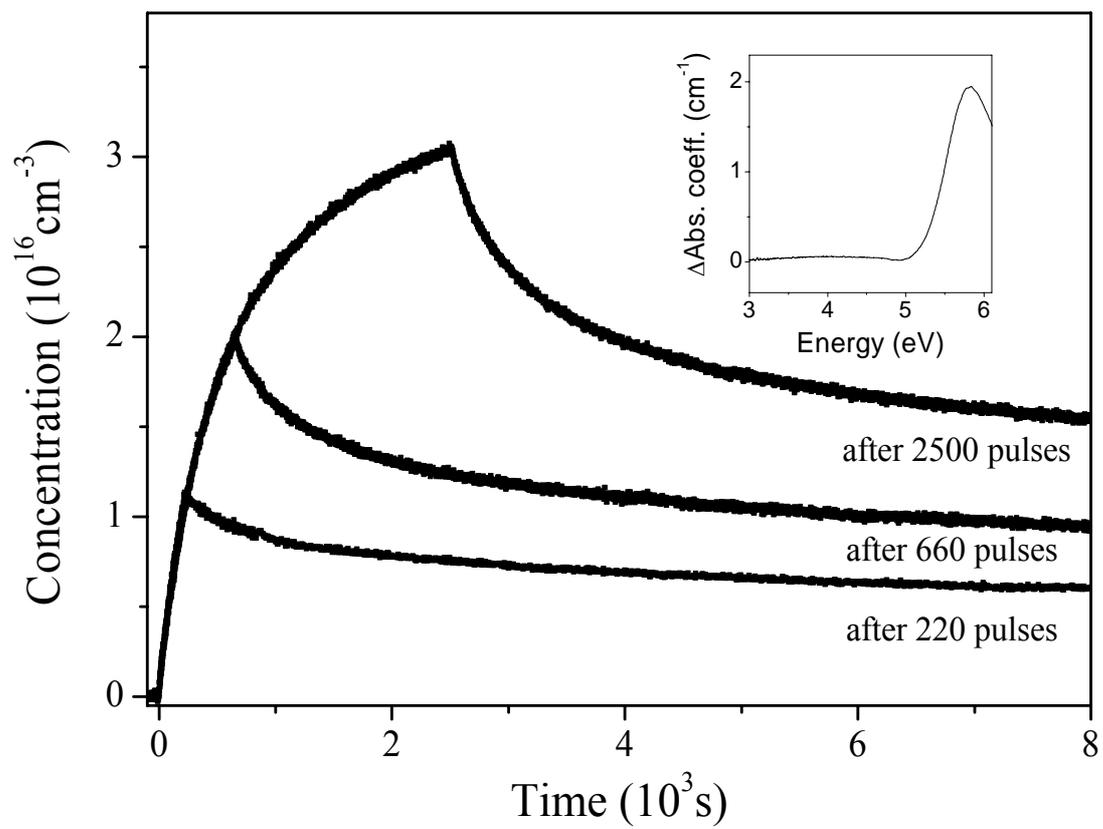

**Figure 1**

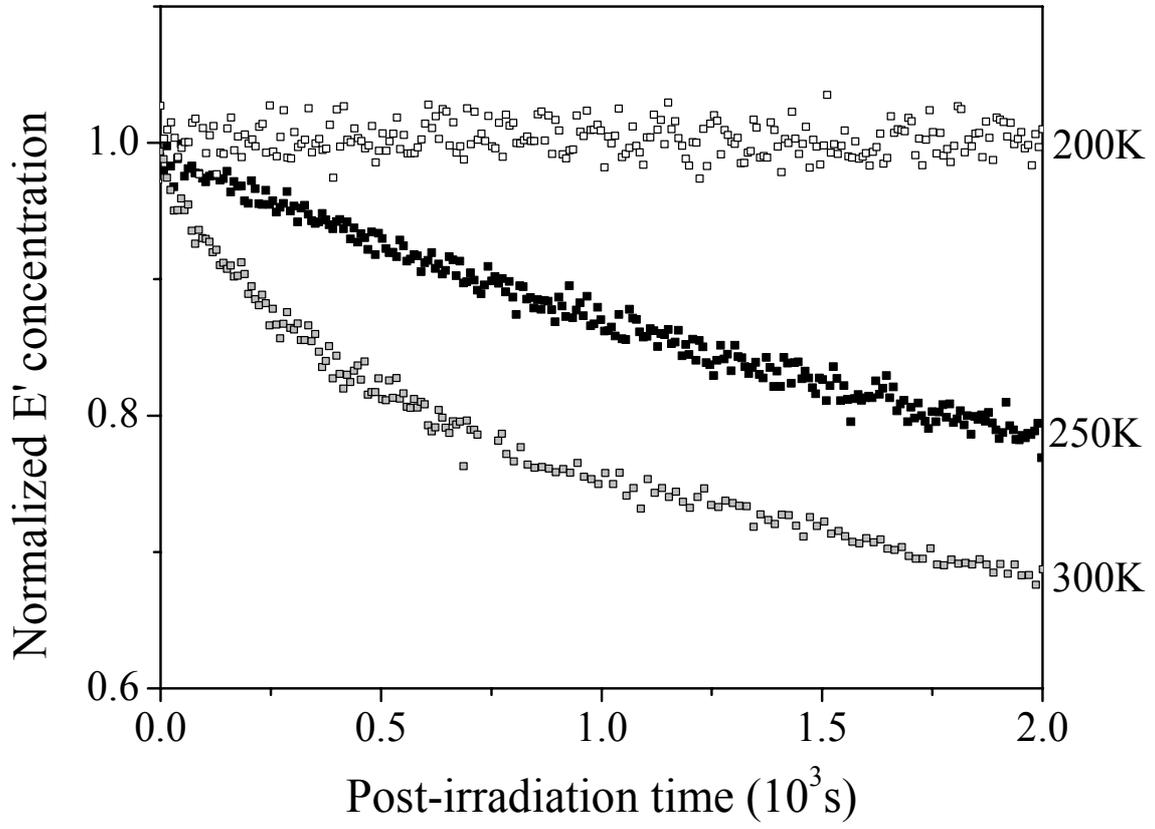

**Figure 2**

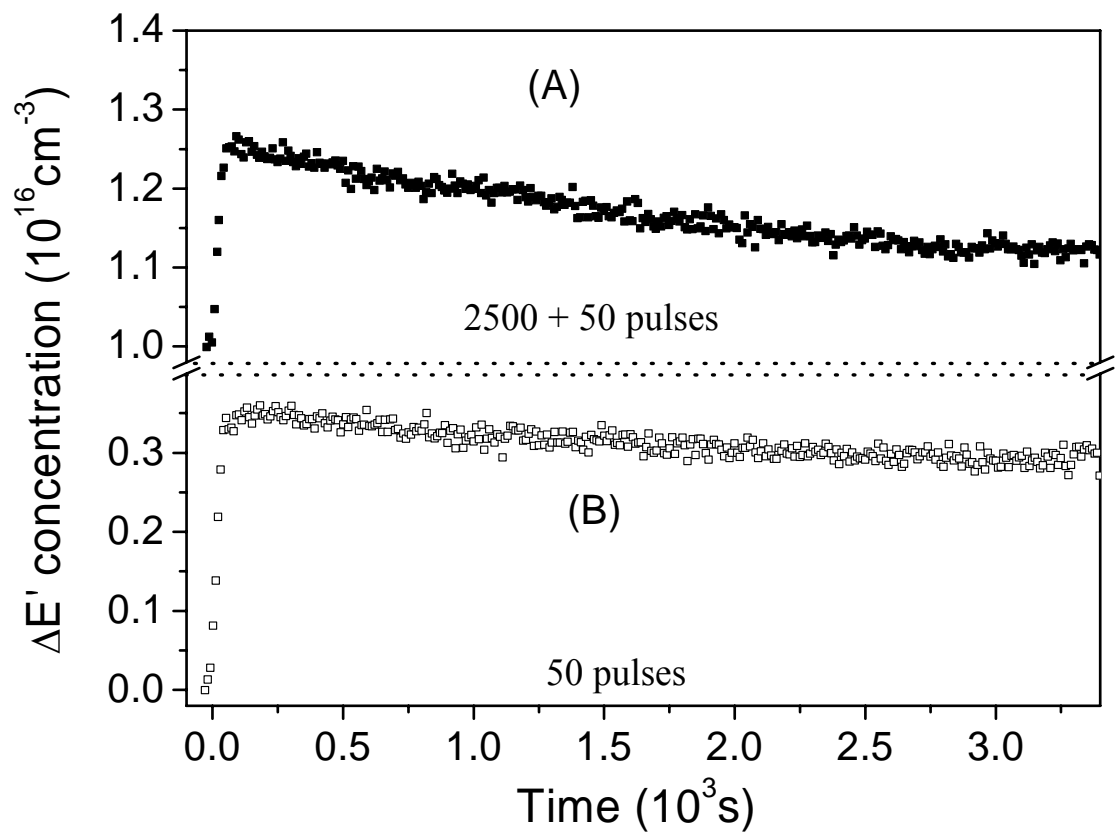

**Figure 3**